\documentclass{iopart}
\usepackage{graphicx}

\begin{document}

\title{
Chi-square test on candidate events from CW signal coherent searches
}

\author{Y. Itoh\dag, M.A. Papa\dag, B. Krishnan\dag $~$and X. Siemens\ddag}
\address{\dag Max Planck Institut f\"ur  
Gravitationsphysik, Albert Einstein Institut, \\ 
Am M\"uhlenberg 1, Golm 14476, Germany
}
\address{\ddag Center for Gravitation and Cosmology, Department of Physics, 
University of Wisconsin - Milwaukee, \\
P.O.Box 413, Wisconsin 53201, USA
}

\ead{yousuke.itoh@aei.mpg.de}

\begin{abstract}

In a blind search for continuous gravitational wave signals scanning a wide frequency band one looks for {\it candidate events} with significantly large values of the detection statistic.
Unfortunately, a noise line in the data may also produce a 
moderately large detection statistic.

In this paper, we describe how we can distinguish between noise line
 events and actual continuous wave (CW) signals, 
based on the shape of the
detection statistic as a function of the signal's frequency. 
We will analyze the case of a particular detection statistic, 
the $F$ statistic, proposed by Jaranowski, Kr\'olak, and Schutz. 

We will show that for a broad-band 10 hour search, with a false
 dismissal rate smaller than $10^{-6}$, our method rejects about $70\%$ of 
the large candidate events found in a typical data set from the second 
science run of the Hanford LIGO interferometer.

\end{abstract}

\maketitle

\section{Introduction}

High power in a narrow frequency band (spectral lines) 
are common features of an interferometric gravitational 
wave (GW) detector's output. Although continuous gravitational 
waves could show up as lines in the frequency domain, given the current sensitivity of GW detectors it 
is most likely that large spectral features are noise of terrestrial 
origin or statistical fluctuations. 

Monochromatic signals of extraterrestrial origin are subject to a Doppler modulation due to the detector's relative motion with respect to the extraterrestrial GW source, while those of terrestrial origin are not. Matched filtering techniques to search 
for a monochromatic signal from a given direction in the sky demodulate the data based on the expected frequency modulation from a source in that particular direction. In general this demodulation procedure decreases the significance of a noise line and enhances that of a real signal. However, if the noise artifact is large enough, even after the demodulation it might still present itself as a statistically significant outlier, thus a candidate event.
Our idea to discriminate between an extraterrestrial signal 
and a noise line is based on the different effect that the demodulation procedure has on a real signal and on a spurious one. 

If the data actually contains a signal, the detection statistic presents a very particular pattern around the signal frequency which, in general, a random noise artifact does not. 
We propose here a chi-square test based on the shape of the detection statistic as a function of the signal frequency and demonstrate its safety and its efficiency. We use the $F$ detection statistic described in \cite{JKS98} and adopt the same notation as \cite{JKS98}.
For applications of the $F$ statistic search on real data, see for 
example \cite{Astone03,Abbott03,Allen04a}.

\section{Method}

\subsection{Summary of the method} 
\label{s:summary}

We consider in this paper a continuous GW signal such as we would expect
from an isolated non-axisymmetric rotating neutron star. 
Following the notation of
\cite{JKS98}, the parameters that describe such signal are its emission
frequency $f_s$, the position in the sky of the source $\vec l_s =
(\alpha_s,\delta_s)$, the amplitude of the signal $h_0$, the inclination
angle $\iota$, the polarization angle $\psi$ and the initial phase of
the signal $2 \Phi_{0}$. 

In the absence of a signal $2F$ follows a $\chi^2$ distribution 
with four degrees of freedom (which will be denoted by $\chi^2_4$). 
In the presence of a signal $2 F$ follows a non-central $\chi^2_4$ 
distribution. 

Given a set of template parameters $(\vec l, f)$, the detection statistic 
$F$ is the likelihood function maximized with respect to the parameters 
$\vec p_s = (h_0,\iota,\psi,\Phi_0 )$. $F$ is constructed by combining 
appropriately the complex amplitudes $F_a$ and $F_b$ representing 
the complex matched filters for the two GW polarizations. And given 
the template parameters and the values of $F_a$ and $F_b$ it is possible 
to derive the maximum likelihood values of 
$(h_0,\iota,\psi,\Phi_0 )$ -- let us refer to these as $\vec p_{MLE}$. 
It is thus possible for every value of the detection statistic to 
estimate the parameters of the signal that have most likely generated it.
So, if we detect a large outlier in $F$ we can estimate the associated 
signal parameters: $(\vec l, f, \vec p _{MLE})$. 
Let us indicate with $\tilde s(t)$ the corresponding signal estimate. 

Let $x(t)$ be the original data set, and define a second data set
\begin{equation}
\tilde x(t) \equiv  x(t) - \tilde s(t)
\end{equation}
If the outlier were actually due to a signal $s(t)$ 
and if $\tilde s(t)$ were a good approximation to $s(t)$, then 
$2 \tilde F$ constructed from $\tilde x(t)$ would be $\chi^2_4$ distributed.

Since filters for different values of $f$ are not orthogonal, 
in the presence of a signal the detection statistic $F$ presents 
some structure also for values of search frequency that 
are not the actual signal frequency. For these other frequencies $2 \tilde F$ 
is also $\chi^2_4$ distributed if $\tilde s(t)$ is a good 
approximation to $s(t)$.

We thus construct the veto statistic ${\cal V}$ by summing the 
values of $2\tilde F$ over more frequencies. In particular we 
sum over all the neighbouring frequency bins that, 
within a certain frequency interval, are above a fixed 
significance threshold. We regard each such collection of frequencies 
as a single ``candidate event'' and assign to it the frequency of 
the bin that has the highest value of the detection statistic. 
The veto statistic is then:
\begin{equation}
{\cal V} := \sum_{k\in {\rm event}} 2 \tilde F(f_k).
\end{equation}

In reality, since our templates lie on a discrete grid, the parameters
of a putative signal will not exactly match any templates' parameters
and the signal estimate $\tilde s(t)$ will not be exactly correct. As a
consequence $\tilde x(t)$ will still contain a residual signal and
$\tilde F$ will not exactly be $\chi^2_4$ distributed. The larger the
signal, the larger the residual signal and the larger the expected value
of ${\cal V}$. Therefore, our veto threshold ${\cal V}_{thr}$ will not
be fixed but will depend on the value of $F$. We will find such
$F$-dependent threshold for ${\cal V}$ based on Monte 
Carlo simulations. The signal-to-noise ratio (SNR) for any given value 
of the detection statistic can be expressed in terms of the detection 
statistic as $\sqrt{2F}$, as per Eq. (79) of \cite{JKS98}. 
Therefore we will talk equivalently of an SNR-dependent or 
$F$-dependent veto threshold.

\subsection{Stationary Gaussian noise plus a signal with exactly known parameters}

Let us first examine the ideal case where the detector output consists of  
stationary random Gaussian noise plus a systematic time series  
(a noise line or a pulsar signal) that produces a candidate in the 
detection statistic $F(f)$ for some template sky position $\vec l$ and 
at frequency $f$. The question that we want to answer is: is the shape 
of $F(f)$ around the frequency of the candidate 
consistent with what we would expect from a signal ? 

Our basic observables are 
the four real inner products $X_i(f, \vec l)$ between the observed time series $x(t)$ and the four filters $h_i(t;\vec l,f)$:
\begin{equation} 
X_i(f, \vec l) = (x(t)~ ||~ h_i(t;\vec l,f)), 
\end{equation}
where $i$ runs from $1$ to $4$. The inner product is defined by Eq.(42) of \cite{JKS98}. The four filters $h_i(t;\vec l,f)$ depend on 
the target frequency $f$ and the target sky location $\vec l = (\alpha,\delta)$.

The hypothesis $H_0$ that we would like to examine is 
\begin{eqnarray}
H_0&:& x(t)=n(t)+ s(t;\vec p_{MLE},\vec l,f),
\end{eqnarray}
where $n(t)$ is the detector noise and 
$s(t;\vec p_{MLE},\vec l,f)=A_i(\vec p_{MLE})h_i(t;\vec l,f)$ 
is the template, which in this case perfectly matches the signal. The parameters 
$\vec p_{MLE}$ are the maximum likelihood estimators of ${h_0,\cos\iota,\psi,\Phi_0}$ derived from the data and the template parameters $\vec l$ and $f$. 
The definitions of the four coefficients $A_i$ 
are given in \cite{JKS98}. 

Given that the template parameters $\vec l, f$ exactly match the parameters of the actual signal, then the waveform $\tilde s(t;\vec p_{MLE},\vec l,f)$ exactly matches the actual signal $s(t)$. In this case the 
four variables $\tilde X_i(f, \vec l)$:

\begin{equation}
\tilde X_i(f, \vec l) = X_i(f, \vec l) - (\tilde s(t)~ || ~ h_i(t;\vec l,f))
\end{equation}
are four correlated random Gaussian variables. 
The paper \cite{JKS98} constructs  
the detection statistic $F$ from the data $X_i(f)$. 
Similarly, we construct ${\cal F}_v(f) := F(f;\tilde X(f))$ from the data $\tilde X_i(f)$. ${\cal F}_v(f)$ is also centrally $\chi^2_4$ distributed in the presence of a signal and perfect signal-template match. We obtain the veto statistic by summing $2{\cal F}_v(f)$ over the different frequencies of the event
\begin{equation}
\label{eq:Nu}
{\cal V}:= \sum_{k=k_1,\cdots,k_{N}\in {\rm event}}2{\cal F}_v(f_k),
\end{equation}
where $N$ is the number of the frequency bins in the event.
If the value of ${\cal V}$ is not consistent with 
a $\chi^2_{4N-4}$ distribution, we reject the hypothesis $H_0$.  
Note that the degrees of freedom of the veto statistic is $4N-4$, as 
we use four data points to infer the four parameters $\vec p_s$.

\subsection{Real noise plus a signal with  parameters mismatched with respect to the template}

In the real analysis the signal parameters $\vec l_s$ will not exactly match the values of one of our templates $\vec l$. As a consequence, $\vec p_{MLE}$ will not match exactly the actual $\vec p_s$ parameters and the frequency where the maximum of the detection statistic occurs, $f_{max}$,  will {\it not} be the actual frequency of the signal $f_s$. However we can still set up a procedure to answer the question: is the shape of the $F$ statistic event consistent with what we would expect from a signal with parameters {\it close to} $\vec l$ ?

Suppose that an event has been identified for a position template $\vec l$ and for a value of the signal frequency $f_{max}$. This is how the veto analysis would proceed:  

\renewcommand{\labelenumi}{\arabic{enumi}.} 
\begin{enumerate}
\item we determine $\vec p_{MLE}$ and $X_i(f_k,\vec l)$ for each $f_k$ of the event. 
\item we generate a veto signal 
$
\tilde s(f_k;\vec p_{MLE},\vec l,f_{max}) 
$ 
and compute the four variables 
$ S_i(f_k, \vec l) = (\tilde s(t;\vec p_{MLE},\vec l,f_{max})~||~h_i(t;\vec l,f_k)).$
\item we construct the variables:\\
$ \tilde X_i(f_k, \vec l) = X_i(f_k, \vec l) - S_i(f_k, \vec l).$
\item using Eq. (\ref{eq:Nu}) 
we compute ${\cal{F}}_v(f)$ and then ${\cal V}$  .
\end{enumerate}

If $\tilde s(f_k;\vec p_{MLE},\vec l,f_{max})$ 
is a good approximation to $s(f_k;\vec p_{s},\vec l_s,f_{s})$, 
then ${\cal V}$ 
follows the $\chi^2_{4N-4}$ distribution.

\subsection{SNR-dependent veto threshold}
\label{sec:implicitassumptions}

As already outlined at the end of section \ref{s:summary}, the 
veto statistic does not in general follow a $\chi^2_{4N-4}$ 
distribution because in general the signal parameters do not 
exactly match the template parameters. Due to this mismatch when 
step 3 is performed in the procedure described in the previous 
section, not all the signal is removed from $X_i$. Consequently 
$\cal{V}$ acquires a non-zero centrality parameter. Since this 
scales as $h_0^2$ in the presence of a signal, the veto statistic 
threshold has to change with the SNR of the candidate event in order 
to keep the false dismissal rate constant for a range of different 
signal strengths. We will thus adopt a SNR-dependent veto 
threshold on our veto statistic ${\cal V}$. We will determine 
the threshold ${\cal V}_{thr}(SNR)$ via Monte-Carlo simulations.

An SNR-dependent threshold in a similar context was first used by the  
TAMA group \cite{Tagoshi2001} who performed SNR-${\cal V}/dof$ 
studies to veto out candidate events in their inspiral waves searches.
See also \cite{Allen04b} for a detailed description of a $\chi^2$
time-frequency test. 
In a context of a resonant bar detector burst search, see 
\cite{Baggio2000}.

\section{Application}

To determine the false dismissal rate, the false alarm rate and 
the threshold equation for the veto statistic, we have performed a set of Monte Carlo simulations on artificial and real noise. We have used 10 hours of fake Gaussian stationary noise and of real science data from the LIGO Hanford 4km interferometer. The results presented here are thus valid for a 10 hour observation time, which is the observation time of the all-sky, wide-band search that we plan to conduct on data from the second science run of the LIGO detectors. 
We do not take into account spin down of pulsars. This may be justified
for the short time length of the data.

As it will be explained below we have injected both signals and spurious
noise artifacts of the type that we observe in the detector output. The parameters of the gravitational waves signals which are injected into the noise are uniformly chosen at random in the following ranges:   
$f_{s}\in [100,500]$ Hz, 
$\alpha_s\in [0,2\pi],\sin\delta_s\in [-1,1]$, 
$\cos\iota \in [-1,1]$,  
$\psi \in [-\pi/2,\pi/2]$, 
$\Phi_{0} \in [0,2\pi]$. 
The strain $h_{0}$ or the amplitude of the model noise line is also
randomly chosen in such a way that the resulting detection statistics
value lie in the range: $\sqrt{50} \le \sqrt{2F_{max}} \le 70$. Below
$2F=50$ the efficiency of the test quickly degrades. We will thus not
apply this veto technique to candidate events with $2F < 50 $. In this
sense, our method is designed to only discard large outliers.

\subsection{Safety test}
\subsubsection{Signals in random Gaussian stationary noise}
\label{s:injections_noise}

We have performed $2\times 10^6$ Monte Carlo simulations. The following steps were executed iteratively 200 times:

\begin{itemize}
\item 
We randomly choose a signal frequency $f_s$ of a simulated gravitational 
wave and then follow the steps below 100 times:
\begin{itemize}
\item we randomly choose a signal sky position $\vec l_s = (\alpha_s, \delta_s)$ and perform the steps below 100 times:
\begin{enumerate}
\item we randomly choose a set of $\vec p_s$ signal parameters and generate   
the 10 hour long data set described above consisting of random Gaussian noise and the fake signal .
\item we randomly displace the sky position template from the signal values by adding a random number uniformly distributed between $\pm$ half the sky positions grid spacing:
$|\alpha - \alpha_s| \le 0.01$ and 
$|\delta - \delta_s| \le 0.01$ (both in radians). 
The grid spacing was estimated numerically and ensures that 
the loss in $F$ due to the signal-template mismatch is less than 
5 $\%$ for $99\%$ of the simulations, for a 10 hour observation.
\item we search for the signal with template values $\vec l$ in a small
      frequency range around $f_s$. This results in the identification
      of an event, defined by a value of $F$ (the highest value of all
      the $F$ values in the event, denoted by $F_{max}$) 
occurring at a frequency $f_{max}$.
Based on this we determine the maximum likelihood estimators $\vec p_{MLE}$.
Also, we compute $X_i(f_k, \vec l)$ for all the frequencies of the event.
\item we generate a veto signal $\tilde s(t;\vec p_{MLE},\vec l,f_{max})$.
\item we compute $S_i(f_k, \vec l)$ for all the frequencies of the event.
\item we construct  
${\cal V}$ from 
$\tilde X_i(f_k, \vec l) = X_i(f_k, \vec l) - S_i(f_k, \vec l)$. 
\end{enumerate} 
\end{itemize} 
\end{itemize}

\begin{figure}\centering \includegraphics[height=8cm]{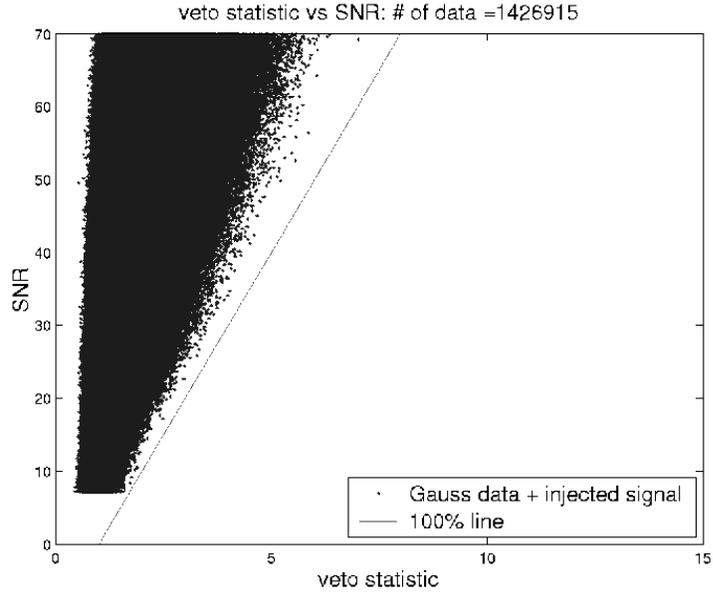}  
\caption{
A scatter plot of  
the veto statistic and the signal to noise ratio (SNR) for 
sets of 10 hours simulated data. 
Each data set consists of 
a Gaussian noise plus a software-simulated signal. 
Each dot in this plot represents 
the candidate event detected by our search code.
The veto statistic in this plot is 
$\tilde {\cal V} \equiv \sqrt{{\cal V}/dof}$, and SNR $=\sqrt{2F}$. 
The straight line represents Eq. (\ref{eq:thresholdline}). 
The detector is assumed to be LHO detector.
The number of the data points with $\sqrt{50}\le SNR\le 70$ is 
1426915.
}
\label{fig:safetygauss}
\end{figure}

\begin{figure}\centering\includegraphics[height=8cm]{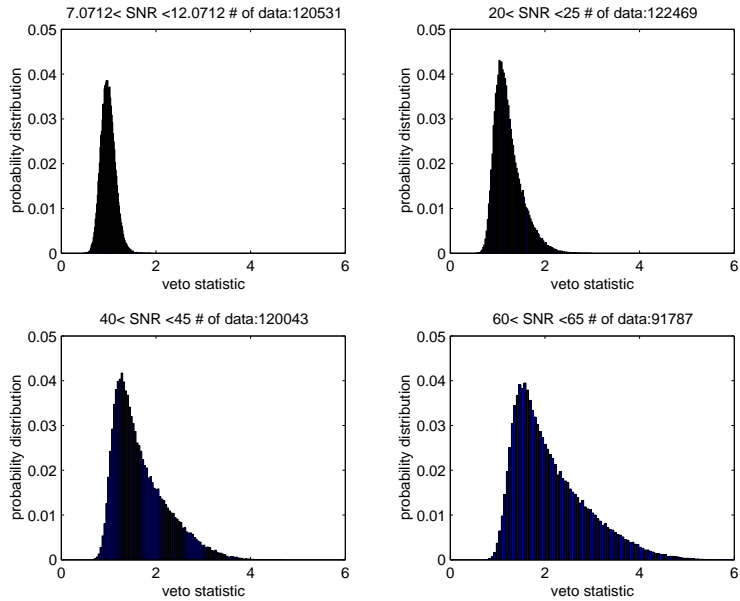} 
\caption{
The estimated probability distributions for $\tilde {\cal V}$ 
with SNR in the four selected ranges.
This figure is for the 10 hours simulated data.
The detector is assumed to be LHO detector.
}
\label{fig:safetygausspdf}
\end{figure}

By considering only values of $\sqrt{50} \le \sqrt{2F} \le 70$ 
we obtain 1426915 sets of $F$ and ${\cal V}$ values. Fig. \ref{fig:safetygauss} is the scatter plot of these. 
It is convenient to normalize each ${\cal V}$ by the 
corresponding number of degrees of freedom, $dof$, since $dof$ could
differ from one injection to another since the number of frequency bins 
in a candidate varies from event to event.  If ${\cal V}$ follows the 
$\chi^2_{dof}$ distribution, then the mean of ${\cal V}$ is just 
$dof$. Fig. \ref{fig:safetygausspdf} shows  
the estimated probability distributions of $\tilde {\cal V} 
\equiv \sqrt{{\cal V}/dof}$ for four selected ranges of $F$.
These four graphs show that the probability distributions 
are well-defined and the Monte Carlo simulations give a good estimate of the 
probability distribution of $\tilde {\cal V}$. Since a variable mismatch exists between the signals and the templates, the distribution of ${\cal V}$ is actually not strictly a central $\chi^2_{dof}$ and the expected value of $\tilde {\cal V}$ is thus not strictly 1. And, as expected, the peak of the distributions of Fig. \ref{fig:safetygausspdf} deviates from 1 more as the signal becomes larger.

From Fig. \ref{fig:safetygauss} one can now define the threshold on the
veto statistic, based on the false dismissal rate that one is willing to
accept. The solid line in the figure, with equation
\begin{eqnarray}
\sqrt{2F_{max}} &=& 10 \sqrt{\frac{{\cal V}}{dof}} - 10 {\rm ~~~(7\times 10^{-7}~~false~~dismissal)},
\label{eq:thresholdline}
\end{eqnarray} 
is the line with the lowest false alarm rate for which, with our sample size, we have not falsely dismissed any of the injected signals. 
In the rest of this paper, 
we will adopt this line, Eq. (\ref{eq:thresholdline}), as the nominal
threshold line.

\subsubsection{Signals in real data -- LHO 10 hours}

We have performed $2\times 10^6$ Monte Carlo simulations by injecting a simulated pulsar signal into real data. 
All the steps are similar to 
those described in \ref{s:injections_noise}. 
We have avoided injections in bands contaminated by spectral disturbances.

From this set of simulations, we obtain a similar 
scatter plot as Fig \ref{fig:safetygauss}. 
And indeed, the threshold line Eq. (\ref{eq:thresholdline}) still 
does not dismiss any injected signals. 

\subsection{Efficiency test}

To study how efficient the test is in vetoing noise artifacts that resemble the signals that we are trying to detect, we have performed an additional set of simulations. For each simulation, 
we have injected sets of time-domain exponentially-damped sinusoids (as a model of a line noise) into both fake Gaussian noise and in real data. In the frequency domain these damped sinusoids have  
a Cauchy distribution, components of which are often observed in the real data.  
We hence follow similar  
steps as for the safety tests described above and produce the corresponding scatter plots of SNR versus $\tilde {\cal V}$.

The efficiency test here 
is ill-defined in the sense that it is possible to generate 
infinite numbers of line noises that have completely different 
shapes from pulsar signals. 
Nonetheless, we think these tests provide ``a feel'' for the efficiency  
of our veto method.

\subsubsection{Noise lines in random Gaussian stationary noise}

The following steps are iteratively performed 200 times:
\begin{itemize}
\item 
we randomly choose the frequency of a noise line and follow the steps below 100 times:
\begin{itemize}
\item we randomly choose a target sky position, with uniform distribution in  
$\alpha\in [0,2\pi],\sin\delta\in [-1,1]$ and then perform the steps below 50 times:
\begin{enumerate}
\item we randomly choose the noise line parameters. The  
e-fold decay rate varies between $0.01/T_0$ and  
$2/T_0$, where $T_0$ is the total observation time.
\item we generate a 10 hour long data set consisting of random Gaussian noise with standard deviation 1 and the noise line defined by the parameters above.
\item we perform a search in a frequency band around the frequency of
      the noise line and identify an event, i.e. a value of $F$ and 
$f_{max}$. From the values of the complex component of the detection
      statistic at $f_{max}$ we determine $\vec p_{MLE}$ and $X_i(f_k, \vec l)$ for every frequency of the event.
\item we generate a veto signal $\tilde s(t;\vec p_{MLE},\vec l, f_{max})$ 
\item we compute $S_i(f_k, \vec l)$  for the veto signal for all the frequencies of the event.
\item we obtain $\tilde {\cal V}$.
\end{enumerate} 
\end{itemize} 
\end{itemize}

\begin{figure}\centering \includegraphics[height=8cm]{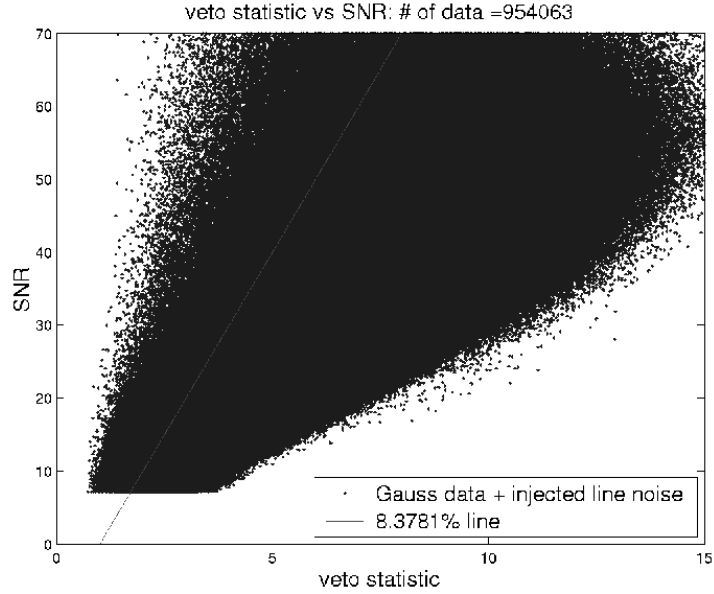}   
\caption{
A scatter plot of  
the veto statistic and the signal to noise ratio (SNR) for 
sets of 10 hours simulated data. 
Each data set consists of 
Gaussian noise plus a software-simulated noise line. 
Each dot in this plot represents 
the candidate event detected by our search code.
The veto statistic in this plot is 
$\tilde {\cal V} \equiv \sqrt{{\cal V}/dof}$, and SNR $=\sqrt{2F}$. 
The straight line represents Eq. (\ref{eq:thresholdline}). 
The detector is assumed to be LHO detector.
The number of the data points with $\sqrt{50}\le SNR\le 70$ is 
954063.
}
\label{fig:efficiencygauss}
\end{figure}

\begin{figure}\centering \includegraphics[height=8cm]{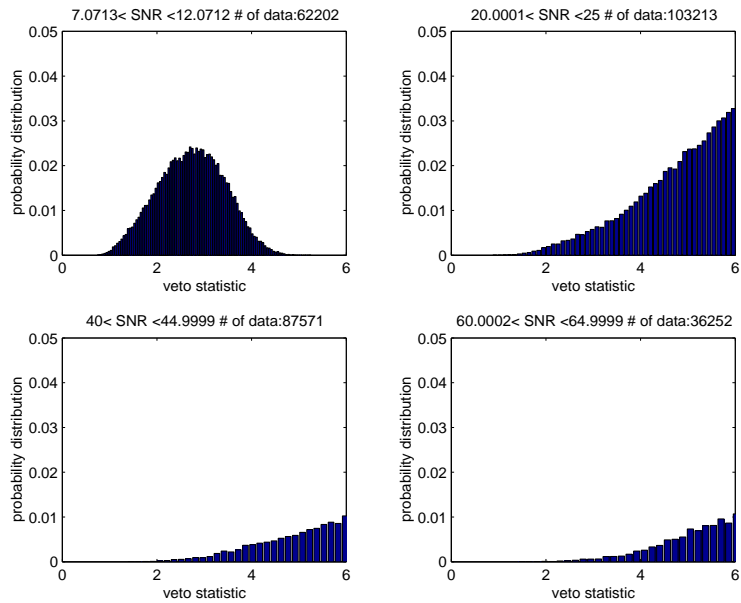}   
\caption{
The estimated probability distributions for $\tilde {\cal V}$ 
with SNR in the four selected ranges corresponding to Fig. 
\ref{fig:efficiencygauss}.
}
\label{fig:efficiencygausspdf}
\end{figure}

Fig. \ref{fig:efficiencygauss} shows the SNR-$\tilde {\cal V}$ 
plot. It may seem that the data points are densely 
distributed in the left upper region with 
large SNRs and small $\tilde {\cal V}$. This deceptive appearance is 
due to the coarse graphical resolution of the figure. 
This can be
clearly seen in the estimated probability distributions, shown in 
Fig. \ref{fig:efficiencygausspdf}. 
In fact, 
if we take our nominal threshold line, Eq. (\ref{eq:thresholdline}), 
shown as the solid line in Fig. \ref{fig:efficiencygauss},    
the false alarm rate is estimated to be 8.4 $\%$.

\subsubsection{Real data: LHO 10 hours}

We have performed $10^6$ Monte-Carlo simulations injecting noise lines as described above into real data, avoiding frequency bands with large noise artifacts. 

The resulting scatter plot is similar to that obtained for the 
Gaussian random noise case. Indeed, we obtain  $5.1 \%$ false 
alarm rate for the nominal threshold Eq. (\ref{eq:thresholdline}).

\subsection{Application to real data}

Having observed safety and efficiency of our veto method, 
we now show an application of the method to real data (no signal nor 
noise lines injected).

We take the following steps iteratively 1200 times: 
\begin{itemize}
\item  we randomly choose a template sky direction $\vec l$ over the whole sky
\begin{enumerate}
\item we perform a wide-band search over the interval [100,500] Hz in the 10 hour real data set.
We identify events in the detection statistic and to each of these events we apply our veto test. This procedure yields a value of $F$ and $\tilde {\cal V}$ for each candidate event.   
\end{enumerate} 
\end{itemize}

\begin{figure}\centering \includegraphics[height=8cm]{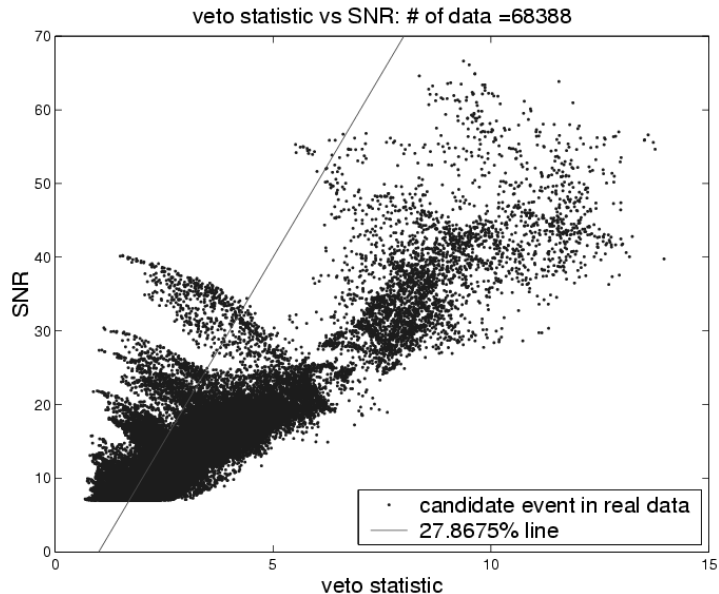}   
\caption{
A scatter plot of  
the veto statistic and the signal to noise ratio (SNR) for 
a 10 hours real data of LHO detector. Each dot in this plot represents 
the candidate event detected by our search code.
The veto statistic in this plot is 
$\tilde {\cal V} \equiv \sqrt{{\cal V}/dof}$, and SNR $=\sqrt{2F}$. 
The straight line represents Eq. (\ref{eq:thresholdline}). 
The number of the data points is 68388. 
The maximum $\sqrt{2F_{max}}$ is 66.6. 
}
\label{fig:realveto}
\end{figure}

The scatter plot SNR-$\tilde {\cal V}$ is shown in Fig. \ref{fig:realveto}. Two distinct branches along the solid line at higher 
SNRs are evident. Both branches are due to spectral features in the data: the highest SNR branch to a line at 465.7 Hz, 
the lower branch to a line at 128.0 Hz. These spectral features ``trigger-off'' a whole set of templates giving rise to the observed structure in the scatter plot.
If we adopt the threshold line Eq. (\ref{eq:thresholdline}), 70 $\%$ of the events are rejected.

\section{Discussions}

We have defined a veto statistic to reject or accept candidate CW events based on a consistency shape test of the measured detection statistic. We have shown how to derive the SNR-dependent threshold for the veto test, through Monte Carlo simulations on a playground data set similar to the one that one intends to analyze. 

\begin{figure}\centering \includegraphics[height=8cm]{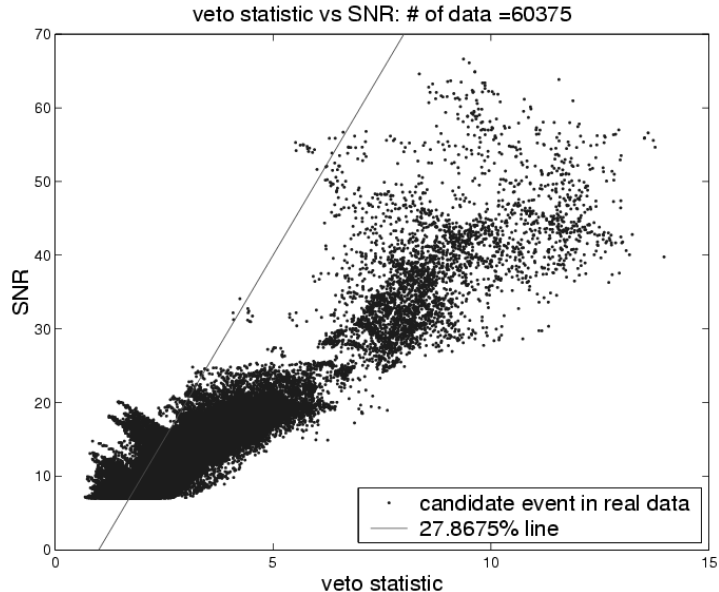}   
\caption{The same plot as \ref{fig:realveto}, but after 
removing the 16 Hz harmonics. 
The number of the data points is 60375. 
The maximum $\sqrt{2F_{max}}$ is 66.6.
}
\label{fig:realvetomike}
\end{figure}

The veto method demonstrated in this paper does not 
require any a-priori information on the source of noise lines. 
However, we expect that the effectiveness of this veto technique can greatly benefit from data characterization studies aimed at identifying spectral contamination of instrumental origin. 
We are now further investigating methods to veto out family of outliers identified in the scatter plots above the solid line in Fig. \ref{fig:realveto}. Natural candidates are those 
noise lines whose properties
are known experimentally, for example the 16 Hz harmonics in the LHO data due to the data acquisition system. It is precisely these harmonics that give rise to one of the major branches above the solid line 
in Fig. \ref{fig:realveto},  as shown in Fig. \ref{fig:realvetomike}.

In this paper, we have used a 10 hour long data set.
For a longer observational time, 
the difference between an extraterrestrial line and a  
terrestrial one becomes larger because the Doppler modulation patterns
of a putative signal carry a more specific signature, that of the motion
of the Earth around the Sun. We have not included spin down of pulsars in our
current study, as we have used short enough time length data. 
Spin down effects of pulsars become more important for a longer 
observation time, and spin down effects generate characteristic feature 
in the $F$ statistic shape. 
We thus expect that our veto method will 
become more  
efficient and safer for longer observation times.  

Finally,  we note that a veto threshold line varies depending on 
observational data time length and noise behavior.  The threshold 
line Eq. (\ref{eq:thresholdline}) is specifically for 10 hours LHO data, 
of particular band, and we recommend that any other search that uses 
quite different data set from our play ground data 
should determine a threshold line based on a play ground data 
in each analysis.

\ack 
We would like to thank M. Landry, who has provided us with the tables of 
the experimentally-measured noise lines 
of the data sets that we have analyzed.  
The work of XS was supported by 
National Science Foundation grants PHY 0071028 and PHY 0079683.

\end{document}